\RequirePackage{fix-cm}
\documentclass[smallextended]{svjour3}       % onecolumn (second format)
\smartqed  % flush right qed marks, e.g. at end of proof
\usepackage{graphicx}
\begin{document}

\title{Interacting dark fluid in the universe bounded by event horizon:A non-equilibrium prescription}



\author{Subenoy Chakraborty         \and
       Atreyee Biswas %etc.
}


\institute{Subenoy Chakraborty  \at
              Department of Mathematics \\
              Jadavpur University\\
              Kolkata-700032
              \email{schakraborty@math.jdvu.ac.in}           %  \\
           \and
          Atreyee Biswas \at
              Department of Natural Science\\
              West Bengal University Of Technology\\
              Kolkata-700064
              \email{atreyee11@gmail.com}
}

\date{Received: date / Accepted: date}

\maketitle

\begin{abstract}
A non-equilibrium thermodynamic analysis has been done for the interacting dark fluid in the universe bounded by the event horizon. From observational evidences it is assumed that at present the matter in the universe is dominated by two dark sectors-dark matter and dark energy. The mutual interaction among them results in spontaneous heat flow between the horizon and the fluid system and the thermal equilibrium will no longer hold. In the present work, the dark matter is chosen in the form of dust while the dark energy is chosen as a perfect fluid with constant equation of state in one case and holographic dark energy model is chosen in the other. Finally, validity of the generalized second law of thermodynamics has been examined in both cases.
\end{abstract}

\section{Introduction}
\label{intro}
At present, based on the recent observational evidences (particularly from Type Ia supernovae observations \cite{Ref1}) it is commonly believed that the matter in the universe is dominated by two dark components - dark matter (about 23\%) and dark energy (above 73\%). Dark matter (DM), the invisible matter without pressure can explain galatic curves and large-scale structure formation while dark energy (DE), an exotic matter with large negative pressure is responsible for the present accelerating phase of the universe. Although there are several proposals for DE candidate, but still the nature of dark energy is completely unknown.\\However, the most natural choice for DE candidate is the cosmological constant ~$\Lambda$~ (having equation of state parameter~ $\omega_{\Lambda}=-1$). Although most of the current data seem to confirm the ~$\Lambda CDM$~ model as a good description of the observed universe, but from theoretical view point this model faces a huge order of discrepancy in the observed value and the theoretically estimated value of $\Lambda$ \cite{Ref2} - there are two well known difficulties namely the "fine tuning" and the "cosmic coincidence" problems and are commonly known as cosmological constant problems \cite{Ref3}.\\
However, there are different candidates for the dynamical DE scenario in the literature to interpret the present accelerating phase of the universe namely a) the quintessence scalar field models \cite{Ref4}, the phantom field \cite{Ref5}, K-essence \cite{Ref6}, tachyon field \cite{Ref7}, quintom \cite{Ref8} etc. b) the DE models including Chaplygin gas \cite{Ref9}, brane world models \cite{Ref10}, holographic and agegraphic Dark energy models \cite{Ref12} and so on.\\
From cosmological view point, it is interesting to consider interactions among the constituent matter components of the universe. But local gravity experiments put strong constraint on the interaction of DE with the baryonic matter \cite{Ref13} while there are no restrictions on the interaction among DE and DM ; rather it is physically reasonable since DE gravitates - it may be accreted by massive compact objects (like BH , neutron star). But this energy flow from DE to DM should be small (but non-zero) from cosmological context.\\
Initially, the coupling between DE and DM was considered to reduce the huge difference between theoretically predicted value and the observed value \cite{Ref14} of the cosmological constant and to solve the coincidence problem \cite{Ref15}. Further it has been shown that a proper choice of the interaction term may influence the perturbation dynamics and affect the lowest multipoles of the CMB spectrum \cite{Ref16,Ref17}. Also recently, the analysis of the supernova data together with CMB and large-scale structure \cite{Ref18} revealed such interaction from expansion history of the universe. Further, in the context of the dynamics of the galaxy clusters, signatures of the interaction between DE and DM has been analyzed  \cite{Ref19}. Moreover, from thermodynamical view point the coupling between DE and DM has been studied \cite{Ref20} considering DE as perfect fluid with a well-defined temperature and it has been shown that at present epoch the energy flow should be from DE to DM for the validity of the second law of thermodynamics \cite{Ref21}. In the present work,we consider the universe containing interacting DE and DM as the matter constituents and it is assumed that the universe bounded by the future event horizon is an isolated thermodynamical system. Due to energy flow between the Dark components the  thermodynamical process is irreversible in nature and as a result the extensive property of the entropy of the whole system will no longer hold. We shall formulate the modified entropy of the whole system and examine the validity of the generalized second law of thermodynamics. The paper is organized as follows : section 2 deals with a general prescription for irreversible thermodynamics with DE as a perfect fluid with constant equation of state while holographic DE model has been studied in section  and 3. At the end in section 4,there is a brief discussion and concluding remarks.

\section{A study of the energy transfer between the dark sectors of the matter distribution:A general thermodynamic prescription}

The metric for homogenous and isotropic FRW model of the universe is given by
\begin{equation}
ds^{2}=h_{ab}dx^{a}dx^{b}+R^{2}d\Omega_{2}^{2}
\end{equation}
where ~$h_{ab}=diag\left(-1,\frac{a^{2}}{1-\kappa r^{2}}\right)$ ~is the metric on the 2-space ~$\left(x^{0}=t,x^{1}=r\right)$,
~ $R=ar$ ~is the area radius and ~$\kappa=0, \pm 1$~ indicate flat, closed and open model of the universe. However, recent observations indicate a closed model with a small positive curvature ~$(\Omega_{k}\approx 0.02)$. Also in the context of the recent observational evidences the present day accelerating universe is dominated by an interacting two fluid system-the dark energy (DE) and the dark matter (DM). The Friedmann equations for FRW metric are
\begin{eqnarray}
H^{2}+\frac{\kappa}{a^{2}}&=& \frac{8\pi G}{3}\left(\rho_{m}+\rho_{d}\right)\\
and~~~~ \dot{H}-\frac{\kappa}{a^{2}} &=& -4\pi G\left(\rho_{m}+\rho_{d}+p_{d}\right)
\end{eqnarray}
where the DE component is a perfect fluid having energy density and thermodynamic pressure ~$\rho_{d}$~ and~ $p_{d}$~ respectively while DM is in the form of dust having energy density ~$\rho_{m}$. Using density parameters namely
\begin{equation}
\Omega_{m}=\frac{8\pi G\rho_{m}}{3H^{2}} , \Omega_{d}=\frac{8\pi G\rho_{d}}{3H^{2}}~~ and~~\Omega_{k}=\frac{\kappa}{a^{2}H^{2}}
\end{equation}
the first Friedmann equation can be written as
\begin{equation}
\Omega_{m}+\Omega_{d}=1+\Omega_{k}
\end{equation}
The energy conservation relations for both the subsystems are
\begin{equation}
\dot{\rho}_{m}+3H\rho_{m}=Q
\end{equation}
and
\begin{equation}
\dot{\rho}_{d}+3H(\rho_{d}+p_{d})=-Q
\end{equation}
The interaction term ~$Q>0$~ indicates an energy flow from DE to DM.The explicit form of Q is chosen in the form \cite{Ref22}
\begin{equation}
Q=3H\lambda \rho_{d}
\end{equation}
with ~$\lambda$, a small dimensionless positive quantity. A positive definite Q is necessary both for the coincidence problem \cite{Ref23} to be solved (or atleast alleviated) \cite{Ref24} and for the validity of the second law of thermodynamics \cite{Ref25}. Probably, the interaction hypothesis was first introduced by Wetterich \cite{Ref14} with the motivation of reducing the extremely large theoretical value of the cosmological constant. Subsequently it was used by Horvat \cite{Ref26} in connection to holography. It is implicitly assumed in most cosmological models that matter and dark energy only couple gravitationally. But it is reasonable to choose the interaction to be zero provided there is some underlying symmetry (still to be discovered). So we should rely on observational evidences. Apparently, the above choice of Q (in equation (8)) looks phenomenological, but different Lagrangians have been proposed in support of it \cite{Ref27}. Note that choice of H in Q is motivated purely by mathematical simplicity as well as from dimensional ground. There is a detailed study of the dynamics of interacting DE models with different choices of Q in ref. \cite{Ref16,Ref28}. However, this phenomenological choice has proven to be compatible with observations like SNIa, CMB, large-scale structure, H(z) and age constraints \cite{Ref29}, and recently in galaxy clusters \cite{Ref30}.\\
We now define the horizons for FRW model.The dynamical apparent horizon, a marginally trapped surface with vanishing expansion, is determined by the relation ~$h^{ab}\partial_{a}R\partial_{b}R=0$, which yields the radius of the apparent horizon as
\begin{equation}
R_{A}=\frac{1}{\sqrt{H^{2}+\kappa/a^{2}}}
\end{equation}
Note that for flat space (i.e ~$\kappa=0$) ~$R_{A}$~ coincides with~ $\frac{1}{H}$, the Hubble horizon. On the otherhand, the radius of the event horizon is characterized by the integral 
\begin{equation}
R_{E}=a\int_{t}^{\infty}\frac{dt}{a(t)}
\end{equation}
and this improper integral exists only for an accelerated expanding universe. Thus although the cosmological event horizon does not exist for all FRW universe, the apparent horizon always do exist and can be considered to be a casual horizon.\\
As we are considering the fluid system in the universe, composed of two subsystems (DE+DM) at different temperatures interacting through exchange of energy, so it is reasonable to employ thermodynamics of irreversible process. Accordingly, starting from the Euler's relation : ~$nTs=\rho +p$~ (n=number density of particles in a comoving volume and s= the entropy per particle), and using the above conservation relations (6), (7) and the conservation relation for number density i.e ~$\frac{\dot{n}}{n}=-3H$~ we have the evolution equations for temperature as
\begin{equation}
\frac{\dot{T}_{m}}{T_{m}}=3H\frac{\lambda}{r}
\end{equation}
and
\begin{equation}
\frac{\dot{T}_{d}}{T_{d}}=-3H(\lambda+\omega_{d})
\end{equation}
where ~$T_{m}$~ and~ $T_{d}$~ are the temperature of the DM component and the DE subsystem respectively,  ~$\omega_{d}~ (-1<\omega_{d}<-\frac{1}{3})$ is the equation of state parameter for the DE and r ($=\frac{\rho_{m}}{\rho_{d}}$) is the ratio of the energy densities of the two subsystems. Thus on integration for constant equation of state parameter  we have
\begin{equation}
T_{m}=T_{0}\left(\frac{r}{r_{0}}\right)\left(\frac{a}{a_{0}}\right)^{-\{2+3(\lambda+\omega_{d})\}}
\end{equation}
and
\begin{equation}
T_{d}=T_{0}\left(\frac{a}{a_{0}}\right)^{-3(\lambda+\omega_{d})}
\end{equation}
where ~$T_{0}$~ is the common temperature of the two subsystems in equilibrium configuration while ~$a_{0},~ r_{0}$~ are the values of the scale factor and the ratio of the energy densities in the equilibrium state. It is to be noted that in deriving equation (13) one has to take into account of the temperature ~$T_{m_{0}}\propto a^{-2}$~ for the DM sector in the absence of interaction.\\
However, in presence of interaction, when the temperature of the system differes from that of the horizon,there will be spontaneous heat flow between  the horizon and the fluid components and hence there will no longer be any thermal equilibrium \cite{Ref22,Ref31,Ref32}. At very early stages  of the evolution of the universe we have ~$T_{m}> T_{d}$ ~and with the expansion of the universe,both the subsystems approach to the equilibrium configuration with common temperature ~$T_{0} $~(when $a=a_{0}$). Subsequently (i.e ~$a> a_{0}$) , the thermal equilibrium is violated due to a continuous transfer of energy from DE to DM with~ $T_{m}<T_{0}<T_{d}$. As we are considering the universe bounded by the event horizon as an isolated system, so at the thermal equilibrium the common temperature~ $T_{0}$~ is nothing but the Hawking temperature at the horizon i.e. ~$T_{0}=\frac{1}{2\pi R_{E}}$~ where ~$R_{E}$~ is the radius of the event horizon for the FRW model.\\
For the present isolated system if we denote the entropies of the two subsystems as~ $S_{m}$ ~and ~$S_{d}$~ and ~$S_{E}$~ is the entropy of the bounding event horizon, then
\begin{equation}
T_{m}\frac{dS_{m}}{dt}=\frac{dQ_{m}}{dt}=\frac{dE_{m}}{dt}
\end{equation}
and
\begin{equation}
T_{d}\frac{dS_{d}}{dt}=\frac{dQ_{d}}{dt}=\frac{dE_{d}}{dt}+p_{d}\frac{dV}{dt}
\end{equation}
while from the Bekenstein area formula,
\begin{equation}
\frac{dS_{E}}{dt}=2\pi R_{E}\dot{R_{E}}
\end{equation}
Here ~$V=\frac{4}{3}\pi R_{E}^{3}$~ is the volume of the universe bounded by the event horizon  and ~$E_{m}=\rho_{m}V$~ and ~$E_{d}=\rho_{d}V$.\\As the overall system is isolated so the heat flow across the horizon ~($Q_{h}$) will satisfy
\begin{equation}
\dot{Q_{h}}=-\left(\dot{Q_{m}}+\dot{Q_{d}}\right)
\end{equation}
In equilibrium configuration, the entropy of the whole system depends on the energy densities and volume only and from the extensive property , it is just the sum of the entropies i.e. ~$S_{m}+S_{d}+S_{E}$. However in non-equilibrium thermodynamics one has to take into account of the irreversible fluxes such as energy transfers in the total entropy and hence the time variation of the total entropy is given by \cite{Ref32,Ref33}
\begin{equation}
\frac{dS_{T}}{dt}=\frac{dS_{m}}{dt}+\frac{dS_{d}}{dt}+\frac{dS_{E}}{dt}-A_{d}\dot{Q_{d}}\ddot{Q_{d}}-A_{h}\dot{Q_{h}}\ddot{Q_{h}}
\end{equation}
where $A_{d}$ and $A_{h}$ are the energy transfer constants between DE and DM within the universe and between the universe and the horizon respectively. Now using equations (15) - (17) the explicit form of different terms on the r.h.s of equation (19) are given by
\begin{eqnarray*}
  \frac{dS_{m}}{dt} &=& -3\pi H^{2}R_{E}^{3}\left(\frac{1+z_{0}}{1+z}\right)^{2+3(\lambda+\omega_{d})}\times\\
  &&\left(\frac{r_{0}}{r}\right)\left\{1+\Omega_{k}-\Omega_{d}(1+\lambda HR_{E})\right\} \\
  \frac{dS_{d}}{dt}  &=& -3\pi H^{2}R_{E}^{3}\left(\frac{1+z_{0}}{1+z}\right)^{3(\lambda+\omega_{d})}\Omega_{d}\times\\
  &&\left(1+\omega_{d}+\lambda HR_{E}\right) \\
  \frac{dS_{E}}{dt} &=& 2\pi R_{E}\left(HR_{E}-1\right)\\
 A_{d}\dot{Q_{d}}\ddot{Q_{d}}&=&-\frac{9}{4}A_{d}H^{4}R_{E}^{3}\left(1+\omega_{d}+\lambda HR_{E}\right)\left[2\left(1+\omega_{d}\right)\right.\\
 && \left.+HR_{E}\{3\left(1+\omega_{d}\right)^{2}+\left(3\lambda-2\right)(1+\omega_{d})+3\lambda\}\right.\\
 &&\left.+H^{2}R_{E}^{2}\{3\lambda^{2}+\lambda\left(3\omega_{d}+q+1\right)\}\right]\\
 A_{h}\dot{Q_{h}}\ddot{Q_{h}}&=&-\frac{9}{4}A_{d}H^{4}R_{E}^{3}\left(1+\Omega_{k}+\omega_{d}\Omega_{d}\right)\left[2\left(1+\Omega_{k}\right.\right.\\
 &&\left.\left.+\omega_{d}\Omega_{d}\right)
 +2qHR_{E}+\omega_{d}\Omega_{d}HR_{E}\{3(1+\omega_{d})\right.\\
 &&\left.+3\lambda-2\}
 \right]
\end{eqnarray*}
with z, the usual red-shift parameter.\\
As the expression for ~$\frac{dS_{T}}{dt}$~ is very lengthy, so to get an idea about its sign we make use of the observed or estimated values of different parameters present in the above expressions at present epoch (i.e z=0) as follows \cite{Ref32,Ref33}:\\\\
$\omega_{d}=-1,~~ \lambda=\frac{1}{3},~~ z_{0}=5.56\times10^{7},~ ~r_{0}=1.09\times 10^{5},~ ~\Omega_{d}=0.72$,~~~
$\Omega_{k}=0.02,~ z=0,~~ q=-0.57$.\\\\
So we have
\begin{eqnarray*}
\frac{dS_{T}}{dt}& =&R_{E}\left[5.9\times10^{5}(HR_{E})^{2}(HR_{E}-1.25)+\right.\\
&&\left.0.2\bar{A_{d}}(HR_{E})^{4}(1.92-HR_{E})+0.28\bar{A_{h}}(HR_{E})^{2}(1.43-HR_{E})\right.\\
&&\left.+6.28(HR_{E}-1)\right]
\end{eqnarray*}
where ~$\bar{A_{d}}=A_{d}H^{2}$~ and ~$\bar{A_{h}}=A_{h}H^{2}$\\\\
So if we take ~$\bar{A_{d}},~ \bar{A_{h}}>0$~ and~ $1.25 R_{A}< R_{E}< 1.43 R_{A}$~ we see that ~$\frac{dS_{T}}{dt}>0$~ i.e generalized second law of thermodynamics (GSLT) holds on the event horizon for the present irreversible thermodynamical system, provided ~$R_{E}>R_{A}$~ and ~$\frac{R_{E}}{R_{A}}$~ is restricted to (1.25,1.43).

\section{Holographic Dark Energy Model}
A typical dark energy model which satisfies the holographic principle is known as holographic dark energy (HDE)model. According to this model using effective quantum field theory the energy density is given by \cite{Ref11}~~$\rho_{d}=\frac{3c^{2}}{R_{E}^{2}}$\\
where~ `c'~ is a dimensionless parameter which may be estimated from observation \cite{Ref11,Ref34} and the radius of the event horizon is chosen as the IR cut-off length to obtain correct equation of state and the desired accelerating universe \cite{Ref11}. So one can write ~$R_{E}$~ as~~~$R_{E}=\frac{c}{\sqrt{\Omega_{d}}H}$\\\\
where~ $\Omega_{d}=\frac{8\pi\rho_{d}}{3H^{2}}$~ is the density parameter.\\\\
At first for simplicity of calculations we use the dark sector as the non-interacting two subsystems namely the HDE and the DM. Then the density parameter evolves as \cite{Ref16}
\begin{equation}
\Omega_{d}'=\Omega_{d}(1-\Omega_{d})\left(1+\frac{2\sqrt{\Omega_{d}}}{c}\right)
\end{equation}
and the variable equation of state parameter for the HDE is
\begin{equation}
\omega_{d}=-\frac{1}{3}-\frac{2\sqrt{\Omega_{d}}}{3c}
\end{equation}
where ~'$\prime$'~ stands for differentiation with respect to x=lna.\\In thermodynamics, starting from Euler's relation, the temperature of the HDE (a perfect fluid with variable equation of state) can be written as
\begin{equation}
T_{d}=T_{d0}(1+\omega_{d})e^{-3\int \omega_{d}dx}
\end{equation}
which on integration using (13) gives
\begin{eqnarray}
T_{d}=T_{d0}\frac{a}{a_{0}}(1-\frac{\Omega_{d}}{c})(1-\sqrt{\Omega_{d}})^{\frac{2}{c+2}}
(1+\sqrt{\Omega_{d}})^{\frac{2}{c-2}}\nonumber\\
\times(1+\frac{2}{c}\sqrt{\Omega_{d}})^{\frac{8}{4-c^{2}}}
\end{eqnarray}

Here ~$T_{d0}$ ~is an integration constant and ~$a_{0}$~ is the value of a when DE and DM are in thermal equilibrium. Also the temperature of the DM subsystem (behaving as dust) varies as the reciprocal of the square of the scale factor \cite{Ref21} i.e\\

\begin{eqnarray}
&&T_{m}\propto T_{m0}a^{-2}\\
or,
T_{m}&=&T_{m0}\left(\frac{a}{a_{0}}\right)^{-2}
\end{eqnarray}
Then by virtue of the extensive property, the entropy of the whole system is just the sum of the entropies of the subsystems and the entropy of the horizon, i.e,
\begin{eqnarray*}
\frac{dS_{T}}{dt}&=&\frac{dS_{m}}{dt}+\frac{dS_{d}}{dt}+\frac{dS_{E}}{dt}
\end{eqnarray*}
Then as before using Gibbs' law to obtain the explicit form of the first two terms on the r.h.s of above equation and using Bekenstein entropy-area formula for the 3rd term we obtain\\
\begin{eqnarray}
\frac{dS_{T}}{dt}&=&2\pi
R_{E}\left[HR_{E}-1-\frac{2}{T_{m0}}\frac{R_{E}\rho_{m}(1+z_{0})^{2}}{(1+z)^{2}}\right.\nonumber\\
&&\left.-\frac{4R_{E}\rho_{d}(1+z)}{3T_{d0}(1+z_{0})}(1-\sqrt{\Omega_{d}})^{\frac{2}{c+2}}(1+\sqrt{\Omega_{d}})^{\frac{2}{c-2}}\right.\nonumber\\
&&\left.(1+\frac{2\sqrt{\Omega_{d}}}{c})^{-\frac{8}{4-c^{2}}}\right]
\end{eqnarray}
Now using $T_{m0}=T_{d0}=\frac{1}{2\pi R_{E}}$, the Hawking temperature associated with\\\\
event horizon when DE and DM are in equilibrium, we obtain
\begin{eqnarray}
\frac{dS_{T}}{dt}&=&2\pi R_{E}\left(x-lx^{2}-1\right)
\end{eqnarray}
 where ~$x=HR_{E}$,~$l=4\pi \left[3\left(1-\Omega_{d}\right)a^{2}+\frac{2 \Omega_{d}b}{a}\right]$,~
 $a=\frac{1+z_{0}}{1+z}$,~\\\\$b=(1-\sqrt{\Omega_{d}})^{\frac{2}{c+2}}(1+\sqrt{\Omega_{d}})^{\frac{2}{c-2}}
(1+\frac{2\sqrt{\Omega_{d}}}{c})^{-\frac{8}{4-c^{2}}}$\\\\
Hence for the validity of GSLT we have ~$l< \frac{1}{4}$~and~ $\alpha R_{A}< R_{E}< \beta R_{A}$~with~$\alpha,~\beta=\left(\frac{1\mp \sqrt{1-4l}}{2l}\right)$. Thus for GSLT to hold,~$\Omega_{d}$~ is restricted and~ $\frac{R_{E}}{R_{A}}$ ~has both upper and lower bound. Note that in this case ~$R_{E}$~ may be less than ~$R_{A}$.\\\\           

Now we shall generalize our model by considering interaction between HDE and DM. The form of the interaction term is chosen same as in the previous section. Then the evolution of the density parameter and the equation of state parameter for HDE are modified as
\begin{eqnarray}
\Omega_{d}'&=&\Omega_{d}(1-\Omega_{d})\left(1+\frac{2\sqrt{\Omega_{d}}}{c}\right)-3\lambda \Omega_{d}^{2}\\
\omega_{d}&=&-(\lambda+\frac{1}{3})-\frac{2\sqrt{\Omega_{d}}}{3c}
\end{eqnarray}
Now integrating the energy conservation equations the explicit form of the energy density components are obtained as:
\begin{eqnarray}
\rho_{m}&=&\rho_{m0}\left(\frac{a}{a_{0}}\right)^{-3}exp\left[3\lambda\int\frac{da}{ar}\right]\nonumber\\
and~~
\rho_{d}&=&\rho_{d0}\left(\frac{a}{a_{0}}\right)^{-2}exp\left[\frac{2}{c}\int \frac{\Omega_{d}^{-1/2}d\Omega_{d}}{(1-\Omega_{d})(1+\frac{2\sqrt{\Omega_{d}}}{c})-3\lambda \Omega_{d}}\right]
\end{eqnarray}
In non equilibrium extended thermodynamics due to irreversible fluxes like energy transfers the entropy of the whole system does not satisfy the extensive property (as in equilibrium case), rather it modifies as equation (19).\\The temperature of the two dark sectors are now given by
\begin{eqnarray}
T_{d}&=&T_{d0}\left(\frac{2}{3}-\lambda-\frac{2\sqrt{\Omega_{d}}}{3c}\right)\frac{a}{a_{0}}I\nonumber\\
and~~~~
T_{m}&=& T_{m0}\frac{r}{r_{0}}\frac{a}{a_{0}}I
\end{eqnarray}
where ~$r=\frac{\rho_{m}}{\rho_{d}}$,$r_{0}=\frac{\rho_{m0}}{\rho_{d0}}$~ and~ $T_{m0}$ ~and ~$T_{d0}$~ are integration constants and \\\\              $I=exp\left[\frac{2}{c}\int \frac{\Omega_{d}^{-1/2}d\Omega_{d}}{(1-\Omega_{d})(1+\frac{2\sqrt{\Omega_{d}}}{c})-3\lambda \Omega_{d}}\right]$ \\\\

However if we restrict ourselves to flat universe,then
\begin{eqnarray}
\frac{\dot{Q_{m}}}{T_{m}}&=&\frac{3\pi c^{3}\sqrt{\Omega_{d}}r_{0}(1+z)H^{-1}}{(1-\Omega_{d})(1+z_{0})I}\left[1+\frac{\lambda c}{\sqrt{\Omega_{d}}}-\frac{1}{\Omega_{d}}\right]\nonumber\\
\frac{\dot{Q_{d}}}{T_{d}}&=&\frac{3\pi c^{3}(1+z)H^{-1}}{\sqrt{\Omega_{d}}\left(\frac{2}{3}-\lambda-\frac{2\sqrt{\Omega_{d}}}{3c}\right)(1+z_{0})I}\left[\frac{2}{3c}\sqrt{\Omega_{d}}-\frac{\lambda c}{\sqrt{\Omega_{d}}}+\lambda-\frac{2}{3}\right]\nonumber\\
\dot{Q_{d}}\ddot{Q_{d}}&=&\frac{\Omega_{d}'}{\Omega_{d}^{2}}\frac{9c^{4}H}{4}\left[\frac{2\Omega_{d}^{2}}{9c^{2}}+\left(\lambda-\frac{2}{3}\right)\frac{\Omega_{d}^{\frac{3}{2}}}{3c}+
\frac{\lambda c\sqrt{\Omega_{d}}(\lambda-\frac{2}{3})}{2}-\frac{\lambda^{2}c^{2}}{2}\right]\nonumber\\
\dot{Q_{h}}\ddot{Q_{h}}&=&\frac{\Omega_{d}'}{\Omega_{d}^{2}}\frac{9c^{4}H}{4}\left[\frac{2\Omega_{d}^{2}}{9c^{2}}+\left(\lambda+\frac{1}{3}\right)\frac{\Omega_{d}^{\frac{3}{2}}}{3c}
+\frac{\sqrt{\Omega_{d}}}{3c}-\frac{1}{\Omega_{d}}+\lambda+\frac{1}{3}\right]\nonumber\\
\frac{dS_{E}}{dt}&=&2\pi R_{E}\dot{R_{E}}=6.28cH^{-1}\left(\frac{c}{\Omega_{d}}-\frac{1}{\sqrt{\Omega_{d}}}\right)
\end{eqnarray}
So the time variation of total entropy reads as

\begin{eqnarray}
\frac{dS_{T}}{dt}&=& \frac{3\pi c^{3}H^{-1}I^{-1}(1+z)}{1+z_{0}}\left[\frac{r_{0}\sqrt{\Omega_{d}}}{1-\Omega_{d}}\left(1+\frac{\lambda c}{\sqrt{\Omega_{d}}}-\frac{1}{\Omega_{d}}\right)\right.\nonumber\\
&&\left.+\frac{\lambda-\frac{2}{3}+\frac{2\sqrt{\Omega_{d}}}{c}-\frac{\lambda c}{\sqrt{\Omega_{d}}}}{\sqrt{\Omega_{d}}\left(\frac{2}{3}-\lambda-\frac{2\sqrt{\Omega_{d}}}{c}\right)}\right]+\frac{9c^{4}\Omega_{d}'H^{-1}}{4\Omega_{d}^{2}}\left[\bar{A_{d}}\left(-\frac{2\Omega_{d}^{2}}{9c^{2}}\right.\right.\nonumber\\
&&\left.\left.-\frac{\left(\lambda-\frac{2}{3}\right)}{9c}\Omega_{d}^{3/2}-\frac{\lambda c(\lambda-\frac{2}{3})}{2}\sqrt{\Omega_{d}}+\frac{\lambda^{2}c^{2}}{2}\right)\right.\nonumber\\
&&\left.+\bar{A_{h}}\left(-\frac{2\Omega_{d}^{2}}{9c^{2}}-\frac{(\lambda+\frac{1}{3})}{9c}\Omega_{d}^{3/2}-
\frac{\sqrt{\Omega_{d}}}{3c}+
\frac{1}{\Omega_{d}}-\lambda-\frac{1}{3}\right)\right]\nonumber\\
&&+6.28cH^{-1}\left(\frac{c}{\Omega_{d}}-\frac{1}{\sqrt{\Omega_{d}}}\right)
\end{eqnarray}
with ~$\bar{A}=AH^{2}$\\
From the above expression for ~$\frac{dS_{T}}{dt}$ ~it is not possible to examine the validity of GSLT, however we give a graphical representation of ~$\frac{dS_{T}}{dt}$~ with the variation of ~$\lambda$~ with fixed energy transfer constants in figure1.Here we have fitted our model with three set of obsreved data namely Plank Data sets \cite{Ref35} as given in the following table:\\
\begin{center}
Table-I
\end{center}
\begin{tabular}{|c|c|c|}
\hline 
\rule[-1ex]{0pt}{2.5ex} Data & c & $\Omega_{d}$ \\ 
\hline\hline 
\rule[-1ex]{0pt}{2.5ex} Plank+WP+SNLS3+Lensing & 0.603 & 0.699 \\ 
\hline 
\rule[-1ex]{0pt}{2.5ex} Plank+WP+BAO+HST+Lensing & 0.495 & 0.745 \\ 
\hline 
\rule[-1ex]{0pt}{2.5ex} Plank+WP+Union 2.1+BAO+HST+Lensing & 0.577 & 0.719 \\ 
\hline 
\end{tabular} 

\begin{figure}[htb]
\begin{minipage}{0.4\textwidth}
\includegraphics[width= 1.0\linewidth]{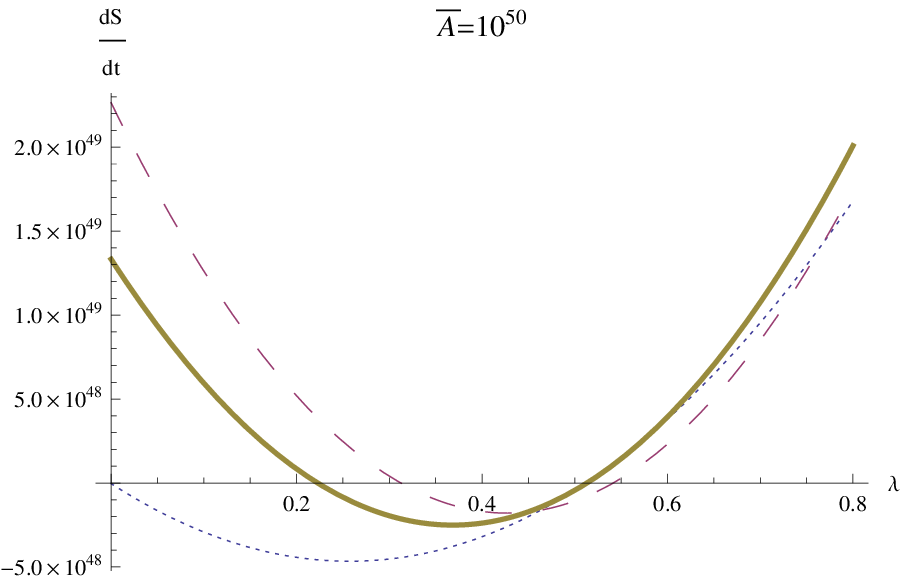}
\end{minipage}
\begin{minipage}{0.4\textwidth}
\includegraphics[width= 1.4\linewidth]{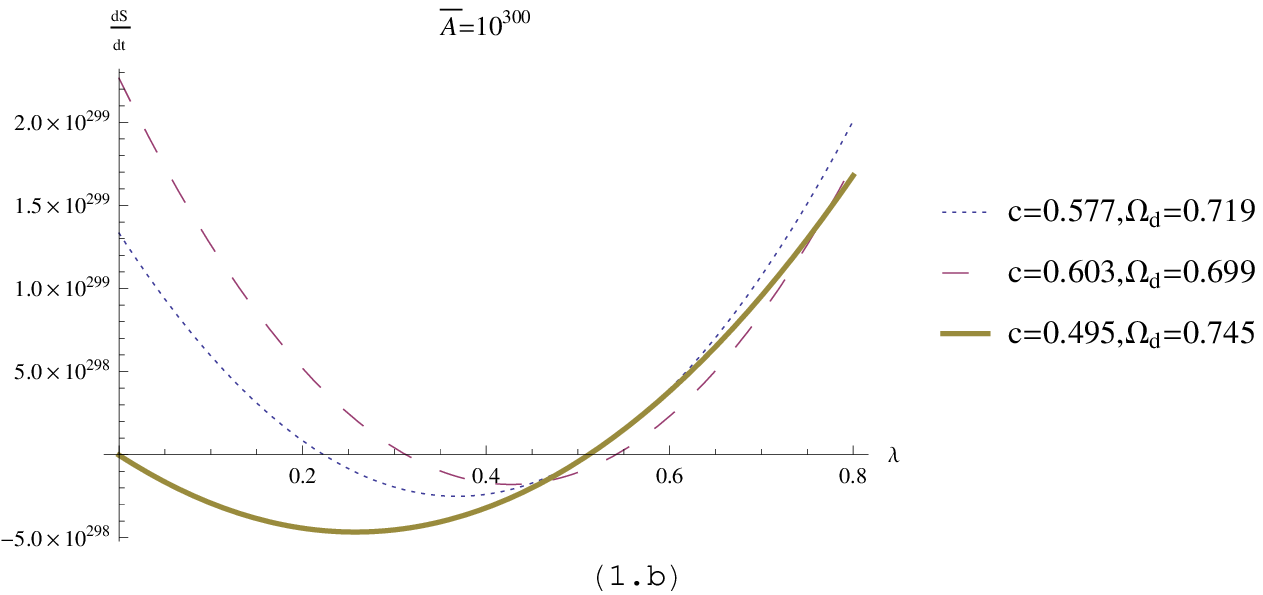}
\end{minipage}
\caption{Graphical representation of~ $\frac{dS_{T}}{dt}$~ in the unit of ~$H^{-1}$~ against ~$\lambda$} ~for ~$z_{0}$~=5.56$\times$ $10^{7}$,~ $r_0$=1.09$\times$ $10^{5}$ ~and (1.a)~$\overline{A}=10^{50}$~(1.b) ~$\overline{A}=10^{300}$
\label{fig:1}
\end{figure}
~\\For simplicity we have assumed ~$\overline{A_{d}}=\overline{A_{h}}=\overline{A}$~ and employed linear approximation to the integral I while evaluating the expressions for temperatures. For fixed  $\overline{A}$ the dependence of ~$\frac{dS_{T}}{dt}$ ~on the parameter $\lambda$ has been shown in figure1 for the present obsrved values of c and ~$\Omega_{d}$~ from TableI. Both the figures (1.a) and (1.b) shows that for approximately ~$0.58<\lambda<1$~ GSLT is always satisfied.
\section{Discussion and Concluding remarks:}

A study of non-equilibrium thermodynamics for the universe bounded by the event horizon has been done with matter content as interacting two-fluid system - the two dark components known at present as dark matter and dark energy.As usual, the dark matter is chosen in the form of dust while in two separate sections the dark energy is known as perfect fluid with constant equation of state and holographic dark energy model respectively. Irreversible thermodynamics is applied to the isolated system (i.e. universe bounded by the event horizon) as the mutual interaction between the two dark fluid species results in a spontaneous heat flow between the horizon and the fluid system. At early epoch of the evolution of the universe the temperature of the DM is larger than that of DE and both approaches the Hawking temperature of the horizon in course of expansion of the universe. However, subsequently this equilibrium configuration is destroyed due to a continuous transfer of energy from DE to DM, and hence the extensive property of the entropy of the whole system can not be applicable to the present system. Though the expression for the time variation of the total entropy of the system is very complicated but it is possible to find restrictions for the validity of the GSLT in case of perfect fluid model of DE and of holographic dark energy model (without interaction). In both cases radius of the event horizon is restricted in a range for which both the bounds are proportional to the radius of the apparent horizon. On the other hand,in case of holographic dark energy interacting with dark matter, even the temperature can only be evaluated in integral form and hence no explicit analytic form for total entropy variation is possible. So we do not have any conclusion regarding validity of GSLT,however,er have shown only graphically.\\Finally, it should be noted that a similar work was done by Karami et al \cite{Ref32} for universe bounded by apparent horizon.But their study was restricted only to DE as perfect fluid with constant equation of state and have shown the validity of the GSLT with a restriction on the energy transfer constants. It is worthy to mention here that in equilibrium thermodynamics GSLT holds unconditionally for universe bounded by apparent horizon, but in non-equilibrium prescription there needs some restriction. On the otherhand, in the present work we have two choices of DE namely i) perfect fluid with constant equation state and ii) holographic dark energy with or without interaction. We have also derived non-equilibrium temperature of the two dark sectors with variable equation of state.Here for validity of GSLT the ratio of two horizon radius are restricted to some range both for perfect fluid with constant equation of state and for HDE without interaction and the restrictions are very similar to equilibrium prescriptions. However for HDE with interaction due to complicated expression we can not derive any analytical restriction, only graphically we have shown the possibile validity of GSLT.

%\begin{figure}
% \includegraphics{atreyee.eps}
%\end{figure}

%\begin{figure*}
 % \includegraphics[width=0.75\textwidth]{example.eps}
%\end{figure*}
%\begin{table}
%\begin{tabular}{lll}
%\hline\noalign{\smallskip}
%\end{tabular}
%\end{table}

\begin{acknowledgements}
The authors are thankful to IUCAA for research facilities as a part of the work is done here.Also they are thankful to the reviewer for his valuable comments.
\end{acknowledgements}


\begin{thebibliography}{}
\bibitem{Ref1}
A.G.Riess et al,Astron.J.,116,1009(1998);\\S.Perlmutter et al,Astrophysics. J.,517,565(1999);\\P.de Bernardis et al,Nature,404,955(2000);\\S.Perlmutter et al.,Astrophysics.J,598,102(2003)\\
\bibitem{Ref2}
S.Weinberg,Rev.Mod.Phys.61,1(1989)\\
\bibitem{Ref3}
E.J.Copeland,M.Sami,S.Tsujikawa,\\Int.J.Mod.Phys.,D15,1753(2006)\\
\bibitem{Ref4}
C.Wetterich,Nucl.Phys.B 302,668(1988);\\
B.Ratra,J.Peebles,Phys.Rev.D 37,321(1988)\\
\bibitem{Ref5}
R.R.Caldwell,Phys.Lett.B 545,23(2002);\\
S.Nojiri,S.D.Odinstov,Phys.562,147(2003);\\
Phys.Lett.B 565,1(2003)\\
\bibitem{Ref6}
T.Chiba,T.Okabe,M.Yamaguchi,Phys.Rev.D 62,023511(2000);\\
C.Armend$\acute{a}$riz-Pic$\acute{o}$n,M.Mukhanov P.J.Steinhardt,Phys.Rev.Lett 85,4438(2000);Phys.Rev.D 63,103510(2001)\\
\bibitem{Ref7}
A.Sen,J.High Energy Phys.,04,048(2002);\\
T.Padmanabhan,T.R.Chaudhury,Phys.Rev.D 66,081301(2002)\\
\bibitem{Ref8}
E.Elizalde,S.Nojiri,S.D.Odinstov,S.Tsujikawa,Phys.Rev.D 71,063004(2005);\\
A.Anisimov,E.Bubichev,A.Vikman,J.Cosmol.Astropart.Phys.,\\06,006(2005)\\
\bibitem{Ref9}
A.Kamenshchik,U.Maschella,V.Pasquier,Phys.Lett.B 511,265(2001);\\
M.C.Bento,O.Bertolami,A.A.Sen,Phys.Rev.D 66,043507(2002)\\
\bibitem{Ref10}
C.Deffayet,G.R.Dvali,G.Gabadadze,Phys.Rev.D,65,044023(2002);\\
V.Sahni,Y.shtanov,J.Cosmol.Astropart.Phys.,11,014(2003)\\
\bibitem{Ref11}
A.Cohen,D.Kaplan,A.Nelson,Phys.Rev.Lett.82,4971(1999);\\
P.Horava,D.Minic,Phys.Rev.Lett.85,1610(2000);\\
S.D.Thomas,Phys.Rev.Lett.89,081301(2002);\\
M.Li,Phys.Lett.B,603,1(2004)
\bibitem{Ref12}
R.G.Cai,Phys.Lett.B 657,228(2007);\\
H.Wei,R.G.Cai,Phys.Lett.B,660,113(2008);\\
Phys.Lett.B 663,1(2008);\\
Eur.Phys.J.C 59,99(2009);\\
K.Y.Kim,H.W.Lee,Y.S.Myung,Phys.Lett.B 660,118(2008);\\
J.Zhang,X.Zhang,H.Liu,Eur.Phys.J.C.54,303(2008)\\
\bibitem{Ref13}
P.J.E.Peebles,B.Ratra,Rev.Mod.Phys.75,559(2003);\\
K.Hagiwara et al;Phys.Rev.D 66,010001(2002)\\
\bibitem{Ref14}
C.Wetterich,Nucl.Phys.B 302,668(1988)\\
\bibitem{Ref15}
L.Amendola,S.Tsujikawa,M.Sami,Phys.Lett.B 632,155(2006);\\
W.Zimdahl,D.Pav$\acute{o}$n,Phys.Rev.D 70,043540(2004);\\
G.Olivares,F.Atrio-Barandela,D.Pav$\acute{o}$n,Phys.Rev.D 74,043521(2006);\\
Phys.Rev.D78,021302(R)(2008)\\
\bibitem{Ref16}
B.Wang,Y.G.Gong,E.Abdalla,Phys.Lett.B 624,141(2005)\\
\bibitem{Ref17}
B.Wang,J.D.Zang,C.-Y.Lin,E.Abdalla,S.Micheletti,Nucl.Phys.B 778,69(2007)\\
\bibitem{Ref18}
C.Feng,B.Wang,Y.G.Gong,R.-K.Su,J.Cosmol.Astropart.Phys.09,005(2007);\\
Z.K.Guo,N.Ohta,S.Tsujikawa,Phys.Rev.D 76,023508(2007);\\
J.H.He,B.Wang,J.Cosmol.Astropart.Phys.06,010(2008);\\
C.Feng,B.Wang,E.Abdalla,R.-K.Su,Phys.Lett.B 665,111(2008)\\
\bibitem{Ref19}
O.Bertolami,F.Gil Pedro,M.Le Delliou,Phys.Lett.B 654,165(2007);\\
E.Abdalla,L.Raul,W.Abramo,L.Sodre,Jr.,B.Wang,arxiv 0710.1198(astro-ph)\\
\bibitem{Ref20}
B.Wang,C.Y.Lin,D.Pav$\acute{o}$n,E.Abdalla,Phys.Lett.B662,19(2008);\\
D.Pav$\acute{o}$n,B.Wang,arXiv:0912.0565\\
\bibitem{Ref21}
H.Callen,Thermodynamics (J.Wiley,1960)\\
\bibitem{Ref22}
D.Pav$\acute{o}$n,W.Zimdahl,Phys.Lett.B 628,206(2005);\\
class Quantum Grav. 24,5461(2007);\\
J.-H.He,B.Wang,J.Cosmol.Astropart.Phys. 06,010(2008)\\
\bibitem{Ref23}
P.J.Steinhardt in:V.L.Fitch,D.R.Marlow(Eds),Critical Problems in Physics,Princeton Univ.Press,Princeton,New Jersey,1997.\\
\bibitem{Ref24}
W.Zimdahl,D.Pav$\acute{o}$n,L.P.Chimento,Phys.Lett.B 521(2001)133;\\
L.P.Chimento,A.S.Jakubi,D.Pav$\acute{o}$n and W.Zimdahl,Phys.Rev.D 67(2003)082513;\\
L.P.Chimento,D.Pav$\acute{o}$n,Phys.Rev.D 73(2006)063511;\\
S del campo,R.Herrera,D.Pav$\acute{o}$n,Phys.Rev.D 74(2006)023501\\
\bibitem{Ref25}
D.Pav$\acute{o}$n,B.Wang,arXiv:0792.0565(gr-qc)\\
\bibitem{Ref26}
R.Harvat,Phys.rev.D 70(2004)087301\\
\bibitem{Ref27}
S.Tsujikawa,M.Sami,Phys.Lett.B 603(2004)113\\
\bibitem{Ref28}
L.Amendold,Phys.Rev.D 60(1999)043501;\\
Phys.Rev.D 62(2000)043511;\\
M.Szydlowski,Phys.Lett.B 632(2006)1;\\
S.Tsujikawa,Phys.Rev.D 73(2006)103504;\\
Z.K.Guo,N.Ohta,S.Tsujikawa,Phys.Rev.D 76(2007)023508.\\
\bibitem{Ref29}
B.Wang,C.Y.Lin,E.Abdalla,Phys.Lett.B 637(2006)357;\\
B.Wang,J.Zang,C.Y.Lin,E.Abdalla,S.Michelelti,Nucl.Phys.B 778(2007)69;\\
C.Feng,B.Wang,Y.Gong,R.-K.Su,J.Cosmol.Astropart.Phys. 09(2007)005\\
\bibitem{Ref30}
O.Bertolami,F.Gil.Pedro,M.Le Delliou,Phys.Lett.B 654(2007)165;\\
Gen.Rel.Grav.41(2009)2839;\\
E.Abdalla,L.R.Abramo,L.sodre,B.Wang,Phys.Lett.B 673(2009)107\\
\bibitem{Ref31}
A.A.Sen,D.Pav$\acute{o}$n,Phys.Lett.B 664,7(2008)\\
\bibitem{Ref32}
K.Karami,S,Ghaffari,Phys.Lett.B 685,115-119(2010)\\
\bibitem{Ref33}
J.Zhou,B.Wang,D.Pav$\acute{o}$n and E.Abdalla,Mod.Phys.Lett.A 24(2009)1689
\bibitem{Ref34}
Q.G.Huang,M.Li,JCAP 0408,013(2004);\\
Z.Chang,F.-Q.Wu,X.Zhang,Phys.Lett.B 633,14(2006);\\
Z.Chang,F.-Q.Wu,Phys.Rev.D 72,043524(2005);\\
Phys.Rev.D 76,023502(2007);\\
E.N.Saridakis,M.R.Setare,Phys.Lett.B 670,01(2008)\\
\bibitem{Ref35}
M Li,X.D.Li,Y.Z.Ma,X.Zhang and Z.Zhang,JCAP 09,021(2013)

\end{thebibliography}


\end{document}